\documentclass[journal,twoside]{IEEEtran}
\usepackage{cite}
\usepackage{amsmath,amssymb,amsfonts}
\usepackage{algorithmic}
\usepackage{graphicx}
\usepackage{textcomp}
\usepackage{subfigure}
\usepackage{booktabs}
\usepackage{makecell}
\usepackage{multirow}

\def\BibTeX{{\rm B\kern-.05em{\sc i\kern-.025em b}\kern-.08em
    T\kern-.1667em\lower.7ex\hbox{E}\kern-.125emX}}
\begin{document}

\title{Deep Reinforcement Learning-Based Battery Conditioning Hierarchical V2G Coordination for Multi-Stakeholder Benefits}
\author{Yubao~Zhang, Xin~Chen*, Yi~Gu, Zhicheng~Li and Wu Kai
   
\thanks{This work was supported in part by the National Natural Science Foundation of China (Grant No.21773182 (B030103)) and the HPC Platform, Xi'an Jiaotong University. (Corresponding author: Xin Chen, e-mail: xin.chen.nj@xjtu.edu.cn)}
\thanks{Yubao~Zhang (e-mail: YubaoZhang@stu.xjtu.edu.cn), Xin Chen (corresponding authour, e-mail: xin.chen.nj@xjtu.edu.cn) and Yi Gu (e-mail: guyi2022@stu.xjtu.edu.cn) are with Center of Nanomaterials for Renewable Energy, State Key Laboratory of Electrical Insulation and Power Equipment, School of Electrical Engineering, Xi'an Jiaotong University, Xi'an, Shaanxi, China.}
\thanks{Zhicheng~Li is with Grid Technology Center, State Grid Fujian Electric Power Research Institute, Fuzhou, Fujian, China (e-mail: li\_zhicheng12@fj.sgcc.com.cn).}
\thanks{Wu Kai is with State Key Laboratory of Electrical Insulation and Power Equipment, School of Electrical Engineering, Xi'an Jiaotong University, Xi'an, Shaanxi, China (e-mail: wukai@mail.xjtu.edu.cn).}
}

\maketitle

\begin{abstract}
With the growing prevalence of electric vehicles (EVs) and advancements in EV electronics, vehicle-to-grid (V2G) techniques and large-scale scheduling strategies have emerged to promote renewable energy utilization and power grid stability. This study proposes a multi-stakeholder hierarchical V2G coordination based on deep reinforcement learning (DRL) and the Proof of Stake algorithm. Furthermore, the multi-stakeholders include the power grid, EV aggregators (EVAs), and users, and the proposed strategy can achieve multi-stakeholder benefits. On the grid side, load fluctuations and renewable energy consumption are considered, while on the EVA side, energy constraints and charging costs are considered. The three critical   battery conditioning parameters of battery SOX are considered on the user side, including state of charge, state of power, and state of health. Compared with four typical baselines, the multi-stakeholder hierarchical coordination strategy can enhance renewable energy consumption, mitigate load fluctuations, meet the energy demands of EVA, and reduce charging costs and battery degradation under realistic operating conditions.

\end{abstract}


\section{Introduction}
\label{sec:introduction}
\subsection{Background and Motivation}
Electric vehicles (EVs) are gaining traction due to several merits of the price reduction and climate and environmental awareness, which neither emit tailpipe pollutants $NO_2$ nor $CO_2$ and have lower maintenance cost and energy cost \cite{sanguesa2021review}. Renewable energy sources (RESs) such as wind and solar can provide energy for EVs without greenhouse gas. Due to the intermittency and unpredictability of renewable energy, energy storage systems should be combined to improve the power quality of renewable energy. The vehicle-to-grid (V2G) is a widely noticed solution, where EVs play a supporting role as a storage resource for the power grid\cite{bibak2021comprehensive}.

\paragraph{V2G Coordination Strategy}

As a type of distributed energy storage, V2G technology has been gaining increasing attention for its potential to improve EV utilization rates and charging/discharging piles\cite{xu2022short}, and regulate frequency\cite{abubakr2023novel}, voltage\cite{hu2021distributed}, and peak shaving\cite{zhang2022transfer} through the implementation of appropriate scheduling strategies. In the V2G coordination strategies, the players, including EVs and power grids, are complex, which brings difficulties and security issues. The intelligent contracts of the V2G trading, two-level optimization strategies were constructed to achieve a fair distribution of benefits among the different market participants, namely EVs, electric vehicle aggregators (EVAs), and the power grid \cite{luo2022hierarchical}. Information sharing between different participants and user privacy protection issues is an important question. For this reason, Luo et al. proposed a vehicle-to-vehicle and V2G electricity trading architecture based on blockchain\cite{luo2021blockchain}. Deep reinforcement learning (DRL) has stronger solving capabilities in unsupervised, nonlinear, and high-dimensional fields than other optimization algorithms or solvers. A DRL technique based on deep Q-networks (DQN) is presented to determine the best charging strategy for EVs, considering the heterogeneity of empirical travel patterns and the unpredictability of electricity prices\cite{hao2023v2g}. As a DRL algorithm for continuous action control, the proximal policy optimization (PPO) algorithm can offer more precise control\cite{schulman2017proximal}. Considering the complex structure of the multi-microgrid systems, a load frequency control strategy for multi-microgrids with V2G based on improved multiagent deep deterministic policy gradient (MADDPG) is proposed\cite{fan2023load}.
\paragraph{Aging Strategy}

The aging of batteries is a significant factor preventing EV customers from becoming participants in V2G. Thus, addressing the battery degradation issue in V2G scheduling is imperative. Wang et al. utilized a long short-term memory neural network to capture the long-term dependencies in the degraded capacities of lithium-ion (Li-ion) batteries\cite{wang2018dynamic}. Fang et al. conducted a deep analysis of battery degradation physics and developed an aging-effect coupling model based on an existing improved single-particle model\cite{fang2023performance}. Augello et al. described a blockchain-based approach for sharing data used both for monitoring the health of the EVs' batteries and for remuneration in V2G for tracking battery usage to permit second-life applications \cite{augello2023certifying}. Optimization models to optimize the charging/discharging schedule with V2G capabilities were proposed, aiming to minimize the charging costs while also considering battery aging, energy costs, and charging costs for the customers \cite{manzolli2022electric, ebrahimi2020stochastic,li2020optimization}. However, these strategies fail to consider the complex relationship between battery aging and state of power (SOP), and state of charge (SOC), which deviates from the realistic operational needs of EVs.
\paragraph{EV Battery Simulation Model}
A battery is the central part of EVs participating in V2G. Different battery states SOX, such as SOC, SOP, and state of health (SOH), play different roles in V2G strategy\cite{shrivastava2023review}. The SOC refers to the energy stored in an EV's battery at a given time. The SOP refers to the ability to provide power over a specific time under the design constraints of EVs' battery. Further, due to concern about potential damage to EV batteries from participating in V2G, participants can use SOH to measure the health of the EVs' batteries. So building the SOX model considering battery SOC, SOP, and SOH is necessary. Accurate linear battery charging models were formulated, closely approximating the real-life battery power and SOC constraints \cite{pandvzic2018accurate, sakti2017enhanced}. In capturing the nonlinearities of Li-ion batteries, Naseri et al. proposed an efficient modeling approach based on the wiener structure to reinforce the capacity of classical equivalent circuit models\cite{naseri2021enhanced}. For the characterization of the steady-state operation of Li-ion batteries, the model characterized the battery performance, including nonlinear charging and discharging power limits and efficiencies, as a function of the SOC and requested power\cite{gonzalez2019non}.

\subsection{Scope and Contributions}

This article proposes a novel battery conditioning hierarchical V2G coordination based on the PPO and Proof of Stake (PoS) algorithms to achieve multi-stakeholder benefits. The multi-stakeholder benefits account for the interests of the power grid, EVA, and users. To schedule large-scale V2G efficiently, the EVA level of the DRL-based coordination strategy considers the EV, EVA, and power grid constraints. Moreover, the lower-level EVA power allocation strategy based on the PoS algorithm considers the EVA power allocation to individual EVs. Then, the PPO and PoS algorithms are adopted to coordinate the large-scale V2G continuous charging/discharging problems. More specifically, our contributions to this article are outlined in the following:
\begin{itemize}
\item First, the proposed multi-stakeholder hierarchical V2G coordination strategy (MHVCS) considers the interests of the power grid, EVAs, and users to achieve multi-stakeholder benefits.
\item Second, PPO and PoS algorithms allow for efficient large-scale V2G coordination, considering grid load fluctuations, renewable energy consumption, energy constraints, charging costs, and battery states. Besides, we add the battery's SOC, SOP, and SOH indicators to the scheduling strategy.
\item Finally, the proposed MHVCS allows for efficient V2G coordination, particularly in large-scale V2G scenarios. Compared with four typical baselines, the proposed strategy can enhance renewable energy consumption, mitigate load fluctuations, meet the energy demands of EVA, reduce charging costs, and reduces battery degradation under realistic operating conditions.
\end{itemize}

\section{Problem Formulation}

\begin{figure*}[htbp]
\centering
\includegraphics[width=0.63\textwidth]{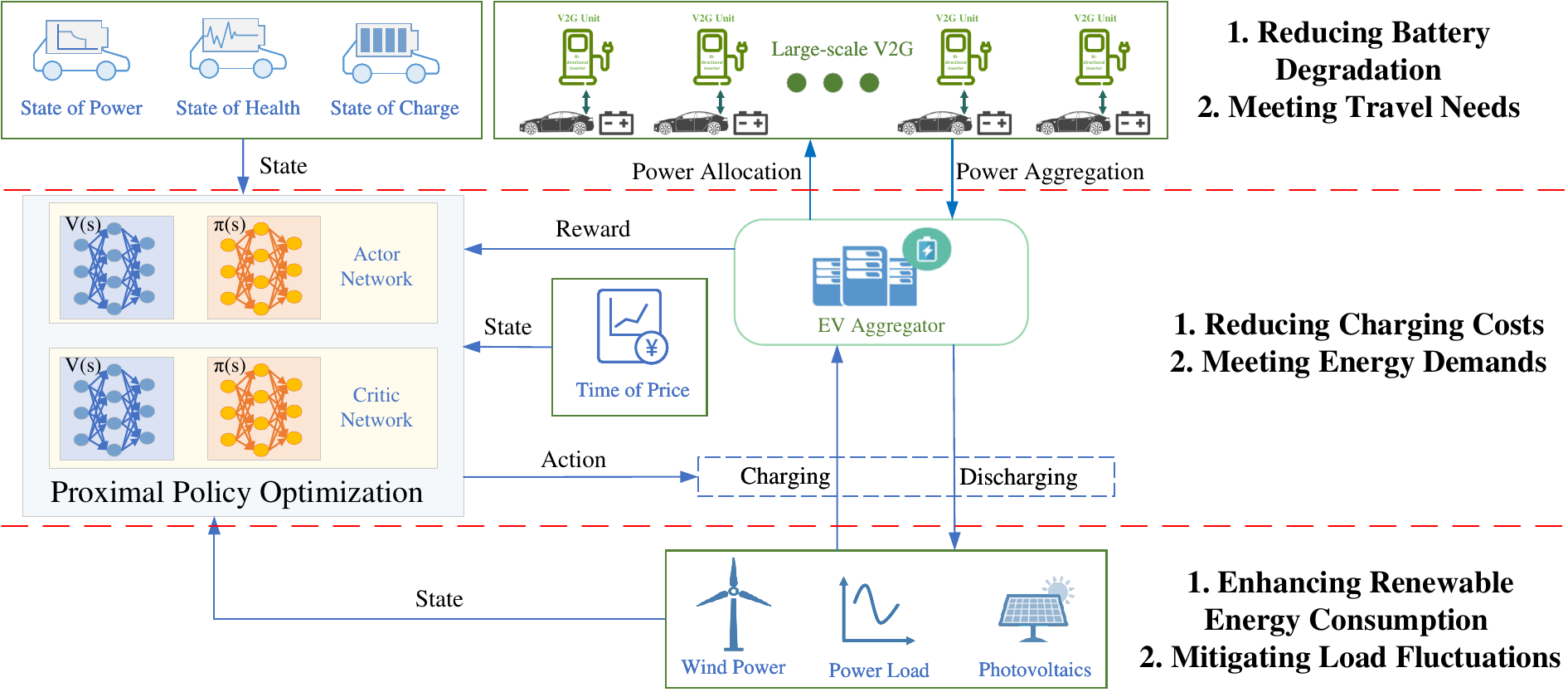}
\caption{Workflow of proposed hierarchical V2G coordination strategy.}
\label{fig:ppoflow}
\end{figure*}

The MHVCS test system comprises wind turbines (WTs), PVs, and EVs, and these distributed energy installations are collectively managed by a central control entity EVA with the PPO algorithm. The MHVCS is presented in Fig.~\ref{fig:ppoflow}.
The DRL-based EVA coordination optimizes the battery states of the examined EVA as an intermediary entity between the EVs and the power grid. The problem is formulated as follows:
\begin{align}
&\max\limits_{P_{EVA}^{k}} F_1+ F_2+ F_3\label{pwgoal}\\
&F_1=\frac{\alpha}{\sigma^{2}}+\beta\left(P_{\max }-P_{\min }\right)+\frac\psi{f_1}\\
&F_2=\left\{\begin{array}{cc}
\chi(\overline{E^k}-E^k)+\upsilon C_{ch} &E^{k}>\overline{E^k} \\
\chi(E^k-\underline{E^k})+\upsilon C_{ch} &E^{k}<\underline{E^k} \\
\end{array}\right.\label{F2}\\
&F_3=\frac \rho {\lvert E^k-E^{k-1} \rvert}
\end{align}
subject to:
\begin{align}
& \sum_{n \in \mathcal{N}} \max \left(\underline{e_n^{k_1}}-\overline{e_n^{k_2}}, \sum_{k_2+1}^{k_1} \underline{p_n^k} \Delta T\right) \nonumber\\
& \quad \leq E^{k_1}-E^{k_2} \leq \nonumber\\
&\sum_{n \in \mathcal{N}} \min \left(\overline{e_n^{k_1}}-\underline{e_n^{k_2}}, \sum_{k_2+1}^{k_1} \overline{p_n^{k}} \Delta T\right)\label{E}\\ 
&\underline{P_{grid}} \leq P^k \leq S*cos\phi\label{ptie}\\
&\underline{P_{EVA}^k} \leq P_{EVA}^k \leq \overline{P_{EVA}^k}\label{peva}
\end{align}
where
\begin{align}
&E^k=E^{k-1}+P_{EVA}^{k}\Delta T\\
&f_1 = \frac1K \sum_{k=1}^{K}P_{load}^k\label{f1}\\
&P_{load}^k = -P_{pv}^k-P_{wt}^k+P_{EVA}^{k}\label{pload}\\
&P^k=P_{base}^k-P_{pv}^k-P_{wt}^k+P_{EVA}^{k}\label{powerload}\\
&\overline{P}=\max{\{P^{k-23},...,P^k\}}\label{maxpower}\\
&\underline{P}=\min{\{P^{k-23},...,P^k\}}\label{minpower}\\
&\sigma^{2}=\frac{1}{K} \sum_{k=1}^{K}\left(P^k-{P}_{ave}\right)^{2}\label{variance}\\
&{P}_{ave}=\frac{1}{K} \sum_{k=1}^{K}P^k\label{ap}\\
&C_{ch}= \sum_{k=1}^Kc^{k}P_{EVA}^{k}\Delta t\label{cost}
\end{align}
The objective function (\ref{pwgoal}) maximizes the overall profit of the power grid $F_1$, EVA $F_2$ and EV $F_3$, which includes the following components: 1) the power grid load variance ${\sigma^{2}}$, the difference between peak load $P_{\max }$ and valley load $P_{\min}$, and the mean net load $f_1$; 2) the EVA energy $E^{k}$ and the charging cost $C$; 3) EV SOH. And the solution variable is ${P_{EVA}^{k}}$.

The variables $\alpha$, $\gamma$, and $\theta$ signify the relative profit weights of the power grid, EVA, and EV users, respectively. $\alpha$ is the reward coefficient of load variance, which is positive. $\beta$ is the penalty coefficient of the peak-valley difference of the power grid, which is negative. $\psi$ is the reward coefficient of the $f_1$, which is negative and ensures the absorption of maximum renewable energy generation. $\chi$ is a positive constraint coefficient of EVA energy, which ensures that the energy does not exceed the boundary value. The positive constraint coefficient of the EVA cost $\upsilon$ ensures minimal EVA cost. $\rho$ is a positive constraint coefficient of the absolute value of EVA energy variation, which ensures minimal depth of discharge for EVs.

Based on the second-order approximation aggregate feasible region \cite{wen2022aggregate}, the EVA energy constraint is $\forall k_1 \in[1, K], \forall k_2 \in$ $\left[0, k_1-1\right]$ shown in (\ref{E}).
$\underline{e_n^{k_1/k_2}}$ is the $n$-th EV minimum energy at timeslot $k_1/k_2$. $\overline{e_n^{k_1/k_2}}$ is the $n$-th EV maximum energy at timeslot $k_1/k_2$. $\underline{p_n^k}$ and $\overline{p_n^k}$ respectively denote the maximum and minimum power of the $n$-th EV at timeslot $k$. $E^k$ is the EVA energy at timeslot $k$. $P_{EVA}^{k}$ is the EVA power at timeslot $k$. $\Delta T$ is the length of a one-time interval. The constraint can be intuitively interpreted as limiting energy change between any two-time points. For example, the right part of the inequality means: that the energy change of the EVA from $k_2$ to $k_1$ does not exceed the energy change from the energy lower bound of $k_2$ to the upper bound of $k_1$ or keeps consuming energy from $k_2$ to $k_1$ with the maximum power. The left part can be explained similarly.

The tie-line power constraint is described in (\ref{ptie}). $\underline{P_{grid}}$ is the minimum tie-line power.
The transformer capacity constraint is described in (\ref{ptie}), where $S$ is the transformer capacity, which is 4000kVA. $cos\phi$ is the power factor, which is 0.8.
The power of EVA constraint is shown in (\ref{peva}) and is subject to the restrictions stipulated by (\ref{E}).
$\overline{P_{EVA}^k}$ and $\underline{P_{EVA}^k}$ are the maximum discharging power and the maximum charging power of the EVA, respectively. $\overline{E^k}$ and $\underline{E}^k$ are the maximum and minimum EVA energy at the timeslot $k$ respectively.
In (\ref{pload}), $P_{load}^k$ is the net load of the microgrid at the timeslot $k$, which is negative. Because the renewable energy generation is negative, the smaller the $P_{load}^k$, the more renewable energy is consumed.
$P^{k}$ is the power load at the timeslot $k$ calculated as (\ref{powerload}).
$\overline{P}$ and $\underline{P}$ are the maximum and minimum power load values, respectively. After scheduling, the calculation is shown in (\ref{maxpower}) and (\ref{minpower}).
The load variance $\sigma^{2}$ indicates the stability of the power grid load. The smaller its value is, the more stable the grid load is. The mathematical model of the load variance is shown as (\ref{variance}), $K$ is the number of timeslots.
$P_{ave}$ is the average of the total power load during the day, which is shown as (\ref{ap}).
$c^{k}$ is the time of use (TOU) tariffs at the timeslot $k$.

\section{PPO and PoS-based MHVCS Approach}
\subsection{PPO-based EVA coordination}
In the DRL algorithms, the agent interacts with an environment through a sequence of observations, actions, and rewards. The agent's goal is to select actions in a fashion that maximizes cumulative future reward\cite{mnih2015human}. DRL can be defined as a Markov decision process that includes: 1) a state space $\mathcal{S}$; 2) an action space $\mathcal{A}$; 3) a transition dynamics distribution with conditional transition probability $p\left(s_{k+1} \mid s_{k}, a_{k}\right)$, satisfying the Markov property, i.e., $p\left(s_{k+1} \mid s_{k}, a_{k}\right)=p\left(s_{k+1} \mid s_{1}, a_{1}, \ldots, s_{k}, a_{k}\right);$ and 4) a reward space $\mathbb{R}$: $\mathcal{S} \times \mathcal{A} \rightarrow \mathbb{R}$.

As a class of continuous DRL algorithms, PPO is applied to the DRL-based EVA coordination strategy for the large-scale EVA that can jointly optimize the grid load. At timeslot $k$, we observe the system state $s_k$. The agent will pick the charging/discharging action $a_k$ based on this information. The action represents the amount of energy the EVA will charge or discharge during the timeslot. After executing the action, we can observe the new system state $s_{k+1}$ and choose the new charging/discharging action $a_{k+1}$ for timeslot $k+1$.

The large-scale V2G continuous charging/discharging coordination problem formulation by the PPO scheduling is presented as Fig.~\ref{fig:ppoflow}. When the EVs park at home, the DRL-based EVA coordination strategy achieves peak shaving, valley filling, and cost reduction by controlling EVA's charging/discharging time and power. Besides, the proposed strategy is also used to reduce the volatility of renewable energy generation.

We detail the DRL formulation of the examined large-scale V2G continuous charging/discharging coordination problem, the critical elements outlined in the following.

\paragraph{Agent} The EVA constitutes the agent, which gradually learns how to improve its retail charging/discharging decisions by utilizing experiences from its repeated interactions with the environment.

\paragraph{Environment} The environment consists of the EVs, power grid, and RES (wind power and PV power), with all of which the EVA interacts.

\paragraph{State}

The state space $\mathcal{S}=\{P^{k-23}, ..., P^k, E^k, {\sigma^{2}_k}, c^k, {\lvert E^k-E^{k-1} \rvert} \}$ encapsulates five types of information:

(1) $P^{k-23}, ..., P^k$ denotes the past 24-hour power grid load values at the timeslot $k$;

(2) $E^k$ represents the EVA energy at the timeslot $k$;

(3) ${\sigma^{2}_k}$ indicates the load variance at the timeslot $k$.

(4) $c^k$ indicates the TOU tariffs at the timeslot $k$.

(5) ${\lvert E^k-E^{k-1} \rvert}$ indicates the EVA energy change at the timeslot $k$.

\paragraph{Action}

Given the state $\mathcal{S}$, $a_{k}$ is the action of an agent at the timeslot $k$.
\begin{align}\label{action}
&\quad a_{k}=P_{EVA}^{k}
\end{align}

The action $a_{k}$ represents the charging/discharging power of the EVA at the timeslot $k$. Let $a_{k}$ be positive when the EVA charges and negative when discharging. Besides, we assume the V2G equipment provides continuous charging/discharging power.

\paragraph{Reward}
The reward is the optimization goal for the agent. To avoid sparse rewards during the agent learning process, the functions below highlight the rewards and penalties for every action undertaken by the agent while scheduling EVA charging/discharging.
\begin{align}
r=F_1+ F_2+ F_3.\label{rr}
\end{align}
where $r$ is the reward, and the goal is to maximize (\ref{rr}).

\subsection{PoS-inspired EVA Power Allocation Algorithm}

With the increasing demand for energy-efficient, secure, and decentralized solutions to manage transactions in smart grids, the PoS algorithm has attracted attention due to its energy efficiency and enhanced security features.

\paragraph{EV User Selection} EV users signal their intention to participate by locking a portion of their battery energy as collateral. The selection process considers factors such as the amount of battery energy, the age of the batteries, and a randomization factor to ensure fair and unbiased participation.

\paragraph{Power Proposal} The selected EV user proposes a new charging/discharging power containing validated transactions. To obtain the individual EV power, the PoS assigns energy weight $\zeta$ to each EV user's battery energy in proportion to the total EV user's battery energy. Thus, the general principle remains constant: EV users with more battery energy are more likely to charge and discharge.

\paragraph{Validation and Consensus} Other EV users in the network verify the proposed charging/discharging's authenticity by checking if the proposer has met the selection criteria and constraints. A consensus mechanism, such as Byzantine Fault Tolerance, ensures that EV users agree on the validity of the proposed charging/discharging before adding it to the strategy.

\paragraph{Rewards and Penalties} EV users receive rewards in the form of cost savings on charging and mitigating battery aging for their contributions to the power grid. PoS encompasses punitive measures against malicious behaviors, such as deviating from the power plan or early EV usage, resulting in partial battery energy loss. The deterrent system discourages validators from dishonestly and preserves the network's security and integrity.

The energy weight $\zeta$ can be defined as:
\begin{align}
&\zeta_n=\left\{\begin{array}{cc}
\frac{(e_{n}^{k}-e_{n}^{k-1})}{\overline{e_{n}^{k}}-e_{n}^{k-1}} &P_{EVA}^{k}>0 \\
\frac{(e_{n}^{k-1}-e_{n}^{k})}{e_{n}^{k-1}-\underline{e_{n}^{k}}} &P_{EVA}^{k}<0
\end{array}\right.\label{kappa}\\
&e_{n}^{k}=e_{n}^{k-1}+{\eta p_{n}^{i,k^{*}}}\Delta t\label{evsoc}
\end{align}

subject to
\begin{align}
&\overline{P_{n,dis}^{i,k}}\leq p_{n}^{i,k^{*}}\leq \underline{P_{n,ch}^{i,k}}\\
&\underline{SOC_{n}^{k}}{Q^i_n}\leq e_{n}^{k}\leq \overline{SOC_{n}^{k}}{Q^i_n}\label{deps}\\
&P_{EVA}^{k}=\sum\limits_{n \in \mathcal{N}}{{\eta p_{n}^{i,k^{*}}}\Delta t}/{Q^i_n}\label{powerall}
\end{align}

$p_{n}^{i,k^{*}}$ represents the charging and discharging power of the $n-{th}$ battery at timeslot $k$ during the $i-{th}$ equivalent full cycle. $\eta$ is the charging and discharging efficiency of the $n-{th}$ battery. $Q^i_n=Q_nSOH^i_n$ is the battery capacity of the $n-th$ EV after the $i-{th}$ equivalent full cycle. $SOH^i_n$ denotes the state of health of the $n$th battery after the $i-{th}$ equivalent full cycle. $\Delta t$ is the length of the timeslot, which is one hour.

$\underline{SOC_{n}^{k}}$ and $\overline{SOC_{n}^{k}}$ are the minimum and maximum SOC for the $n-th$ EV at the timeslot $k$ respectively. In particular, when $\overline{e_{n}^{k}}=e_{n}^{k-1}$ and $P_{EVA}^{k}>0$, $p_{n}^{k^{*}}=0$; when $\underline{e_{n}^{k}}=e_{n}^{k-1}$ and $P_{EVA}^{k}<0$, $p_{n}^{k^{*}}=0$.

\begin{figure}[h]
\centering
\includegraphics[width=0.4\textwidth]{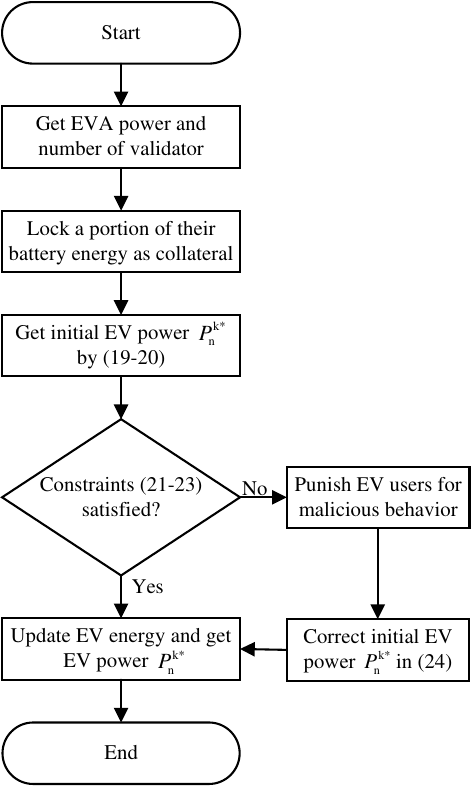}
\caption{Flowchart of the PoS-inspired EVA power allocation algorithm.}
\label{fig:flow}
\end{figure}

After calculating $p_{n}^{k^{*}}$, a safety check correction is needed to obtain the true charging/discharging power $p_{n}^{k}$. During scheduling, the upper and lower limits of the output power of each EV are used to correct the EV energy because the maximum charging and discharging power should not be exceeded. An individual EV should meet the power constraint, and the charging/discharging power $p_{n}^{k}$ is corrected as below.

\begin{align}\label{soc1}
&p_{n}^{k}= \begin{cases}p_{n}^{k^{*}} & \underline{p_{dis}} \leqslant p_{n}^{k^{*}} \leqslant \underline{p_{ch}} \\
\underline{p_{ch}} & P_{EVA}^{k}>0, p_{n}^{k^{*}}>p_{ch}^{max} \\
\underline{p_{dis}} & P_{EVA}^{k}<0, p_{n}^{k^{*}}<p_{dis}^{max}\end{cases}
\end{align}
where $\underline{p_{ch}}$ and $\underline{p_{dis}}$ are the maximum charging and discharging power of the EV, respectively. The EVA power allocation is presented as Fig.~\ref{fig:flow}.

The energy of each EV can effectively converge, and all reach the desired energy before their departure time, solving the redistribution problem caused by the initial energy difference.
\section{Simulation Results}

\subsection{Test System and Implementation}
\begin{figure}[h]
\centering
\includegraphics[width=0.45\textwidth]{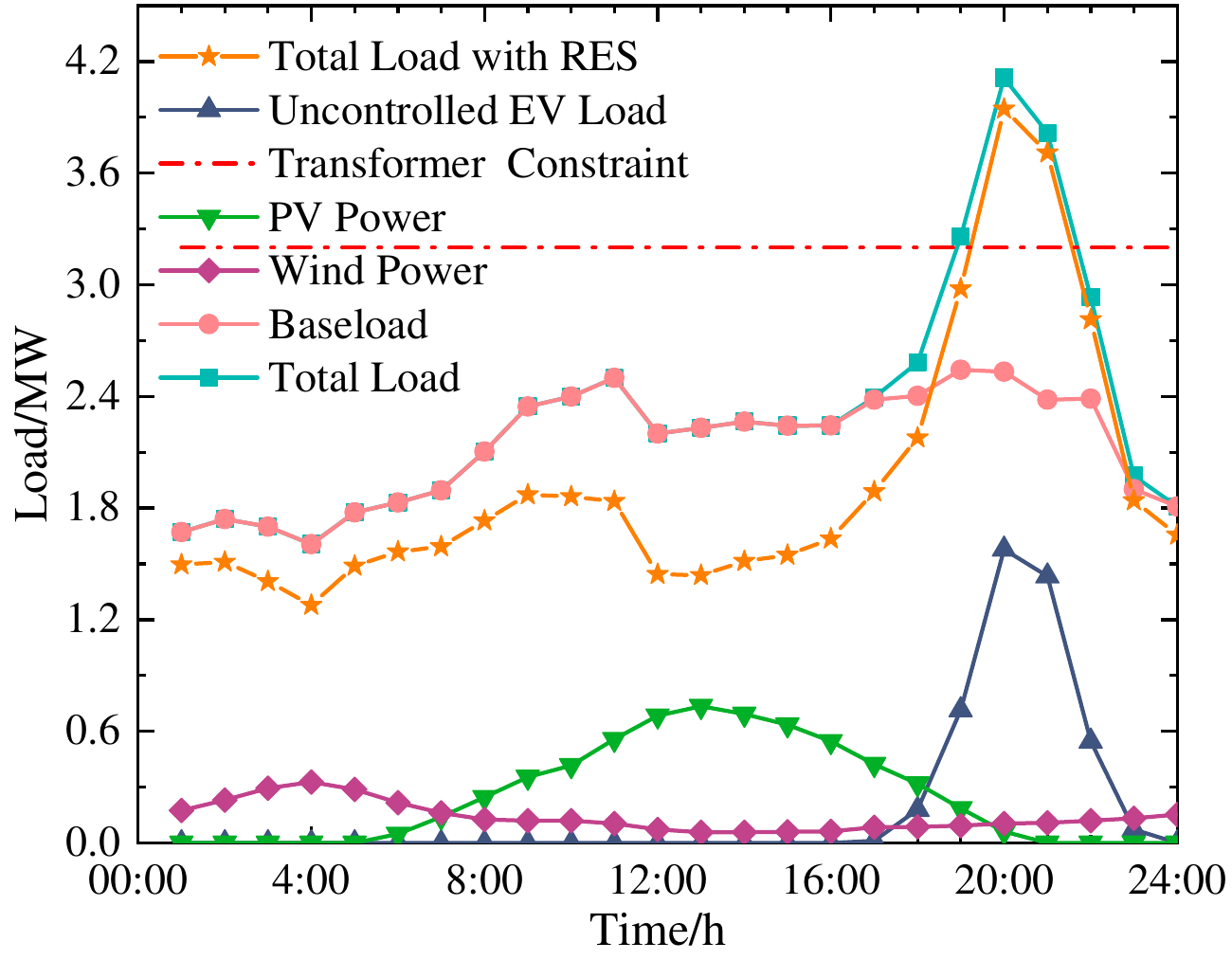}
\caption{Daily load profiles in a microgrid.}
\label{fig:baseload}
\end{figure}

The following simulation results examine the EVA scheduling problem in the context of a single day with hourly resolution. The EVA serves 509 EVs to eliminate EV charging load peaks. The total load, the baseload, and the uncontrolled EV load in the examined day are shown in Figure~\ref{fig:baseload}. The total load profile represents the uncontrolled EV load plus the baseload profile. The total load with and without RES exceed the transformer constraint. With RES, the total load of the microgrid results from the balance of EV, baseload, PV, and wind power.

The assumed values of the remaining technical parameters of the EV are $p_{ch}^{max}= 6 kW$, $p_{dis}^{max}= -6 kW$, $Q_n=24 kWh$ for every EV. The detailed EV load model is shown in \cite{zhang2021optimal}. The arrival time is sampled from $N(18, 1^2)$ and is bounded between 15 and 21. Its distribution $N(8, 1^2)$ for the departure time is bounded between 6 and 10. The SOC of $n-th$ EV $SOC_n$ bounded between 0.2 and 0.8 is sampled from $N(0.5, 0.1^2)$. It should be noted that in the mentioned test systems, the EVA is scheduled from timeslot 15 to 10. To improve the users' participation rate, the enrollment incentive ($\$560$ per customer) is set for users' discharging compensation to compensate users for discharge according to the \cite{9384297}. It is worth noting that the hierarchical coordination strategy does not rely on any knowledge of the distributions of these random variables. Thus, the proposed strategy can be transferred to different modeling mechanisms.

\begin{figure}[h]
    \centering
    \includegraphics[width=0.5\textwidth]{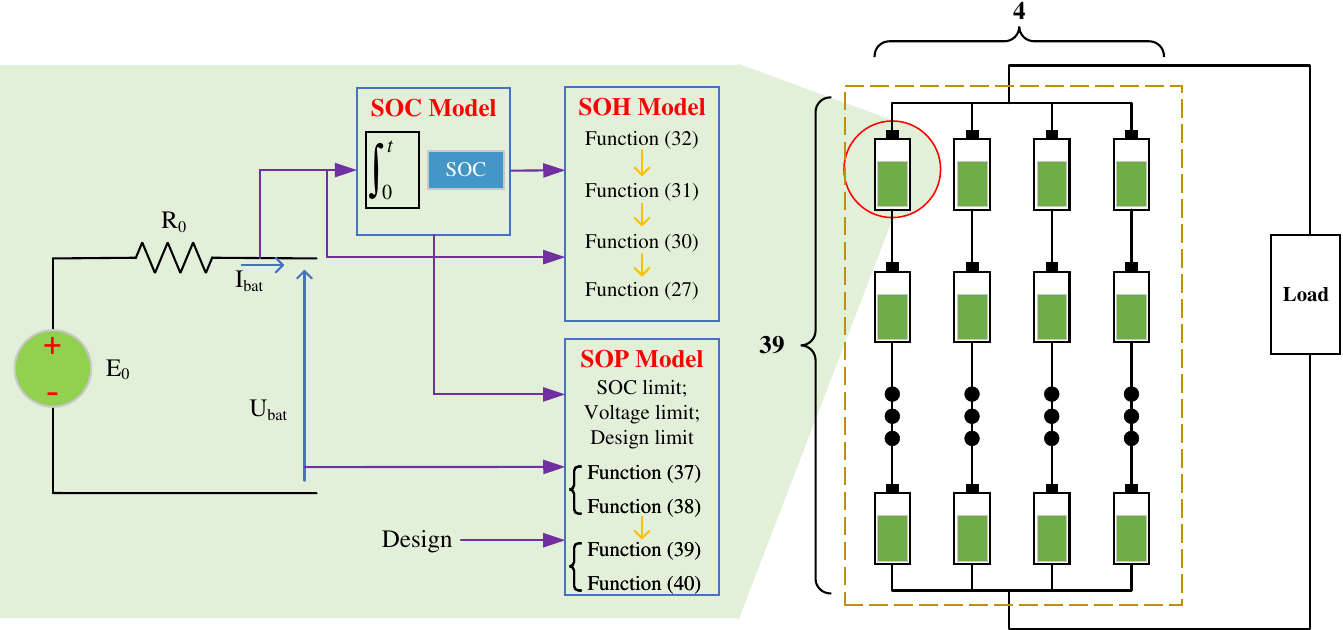}
    \caption{Electric vehicle battery simulation system.}
    \label{fig:simu}
\end{figure}

The topology of the series-parallel battery array is shown in the figure. The model connects battery cells in series and parallel to form a battery module. There are 39 battery cells connected in series and 4 parallel branches. The rated voltage of the battery cell is 3.3V, the rated capacity is 2.3Ah, and the initial equivalent full cycle number of the battery is 50. According to the battery parameters, it is modeled and simulated in MATLAB Simulink as shown in Fig.~\ref{fig:simu}.

\subsection{Training Process of PPO}
The PPO algorithm requires two networks: a critic network for evaluating a state and an actor network mapping a state into a probability distribution over the action space. Both networks are implemented with fully connected layers with the same input size. Both the critic and actor networks have one output neuron and employ the rectified linearity units (ReLU) for all hidden layers. The Adam optimizer is employed for learning the neural network weights with a learning rate $lr^a=10^{-6}$ and $lr^c=2*10^{-6}$ for the actor and critic, respectively. We use a discount factor of $\gamma = 0.95 $ for the critic. We use a hyperparameter $\epsilon=0.2$ to avoid a huge policy update. We use $us = 10 $ as the critic and actor target network updating step. The output layer of the actor is a sigmoid layer to bound the continuous actions. We train with a mini-batch size of $BS=32$ and for $M=3*10^{5}$ episodes, with $L=20$ hours per episode. Finally, we use $\alpha=10$, $\beta={-5}$, $\chi=10$, $\psi=1$, $\upsilon=5$, $\rho=1$ for the penalty weighting constant in (\ref{rr}). The $\alpha, \beta, \psi, \upsilon, \rho, \chi$ values are set manually to ensure that the agent receives the maximum reward and are the result of constant attempts to tune parameters.

\begin{figure}[h]
\centering
\includegraphics[width=0.45\textwidth]{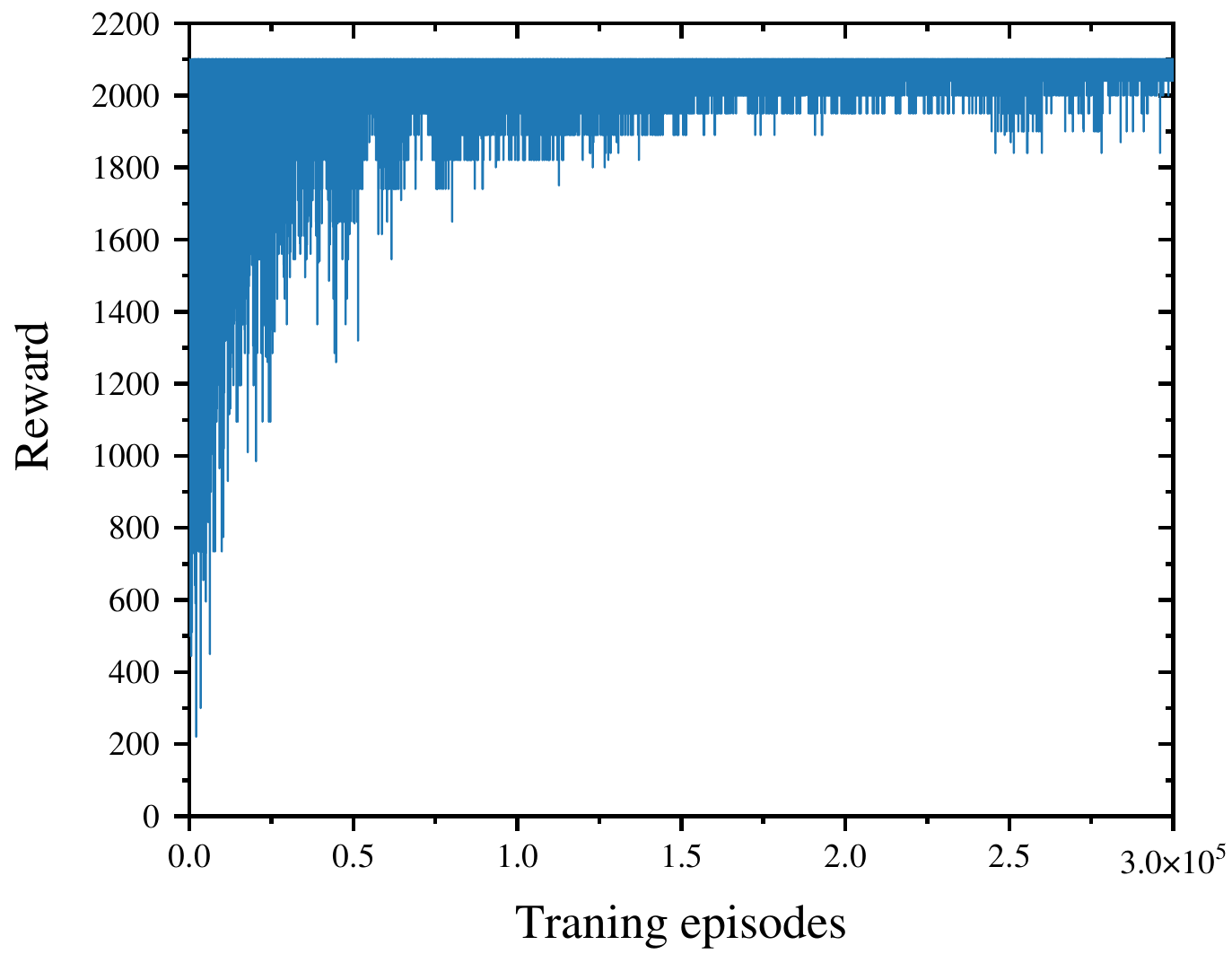}
\caption{The evolution of the rewards during the training process.}
\label{fig:train}
\end{figure}

During the training process, the reward and the load variance over $3*10^{5}$ episodes are calculated and are illustrated in Fig.~\ref{fig:train}.

As demonstrated in Fig.~\ref{fig:train}, the charging/discharging schedule is randomly selected during the initial learning stages since the EVA gathers more experiences by randomly exploring different, not necessarily profitable, actions. However, as the learning process progresses and more experiences are collected, the reward keeps increasing, eventually converging around 2100 with small oscillations. The result demonstrates that the proposed strategy successfully learn a policy to maximize reward.

\subsection{MHVCS Performance}

The performance of the proposed MHVCS is compared with the following baselines:

BL1. Uncontrolled Charging: The user does not consider electricity price and power grid load. When they return home, they charge maximum power until the battery is full.

BL2. Optimal Charging: The user considers battery states and power grid load. They charge according to the EVA's instructions until the battery is full.

BL3. Charging and Discharging with Minimum Grid Load Fluctuation: The user considers grid load level and charges/discharges according to the minimum fluctuation of grid load without considering battery aging until the battery is full.

BL4. Charging and Discharging with Minimum Cost: The user considers TOU tariffs and charges/discharges according to the minimum cost without considering battery aging until the battery is full.

\begin{figure}[h]
\centering
\includegraphics[width=0.45\textwidth]{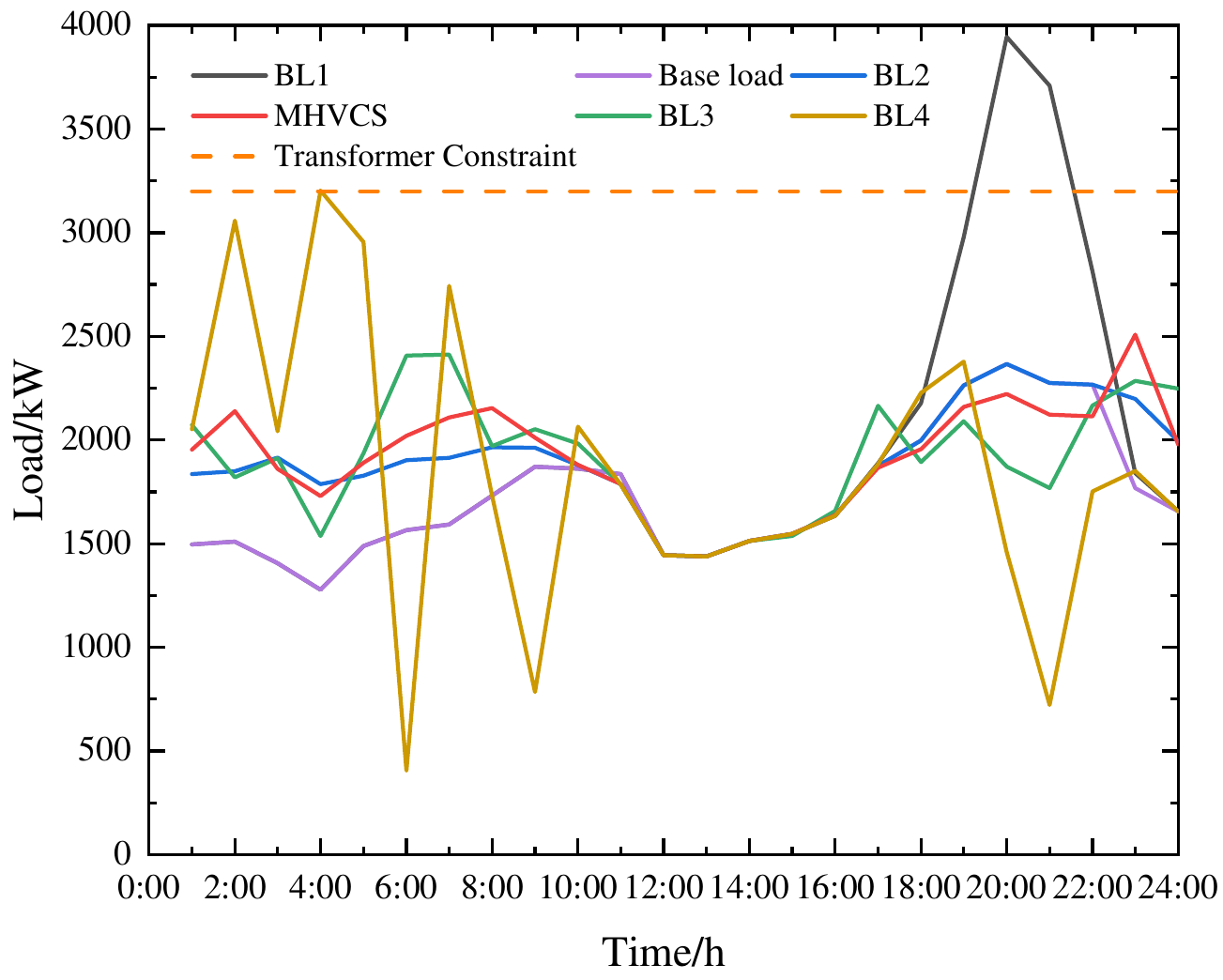}
\caption{Load under MHVCS and four BLs strategies in a day.}
\label{fig:load}
\end{figure}

The performance of the proposed strategy is evaluated with the load variance reflecting the capability of the peak shaving and valley filling during scheduling operation. The load variance is calculated on a test day as (\ref{variance}). The load variance at each timeslot is calculated using the load profile of the previous 24 hours.
In Fig.~\ref{fig:load}, the optimal load profile is the total load profile with EVA scheduled according to the DRL-based EVA coordination strategy. The optimal load profile by PPO is flatter than BL1 with uncontrolled EVA and BL2 with optimal charging. Because compared with the charging mode, V2G can control the discharge to reduce the load peak. The DRL-based EVA coordination strategy by the PPO scheduling achieves peak shaving and valley filling by allocating the EVA charging demand to the load valley.

At the arrival time of timeslot 15, the coordination scheduling starts to connect EVA to the power grid. At the departure time of timeslot 10, the departure load variance becomes 71312.8${kW}^2$. During the scheduling period, the PPO agent gives charging/discharging actions according to the states $\mathcal{S}$. EVA discharges at the peak timeslot 16-21 and is charged at the remaining time in the scheduling period. The SOH under different strategies in one year Fig.~\ref{fig:load}.

\begin{figure}[h]
\centering
\includegraphics[width=0.45\textwidth]{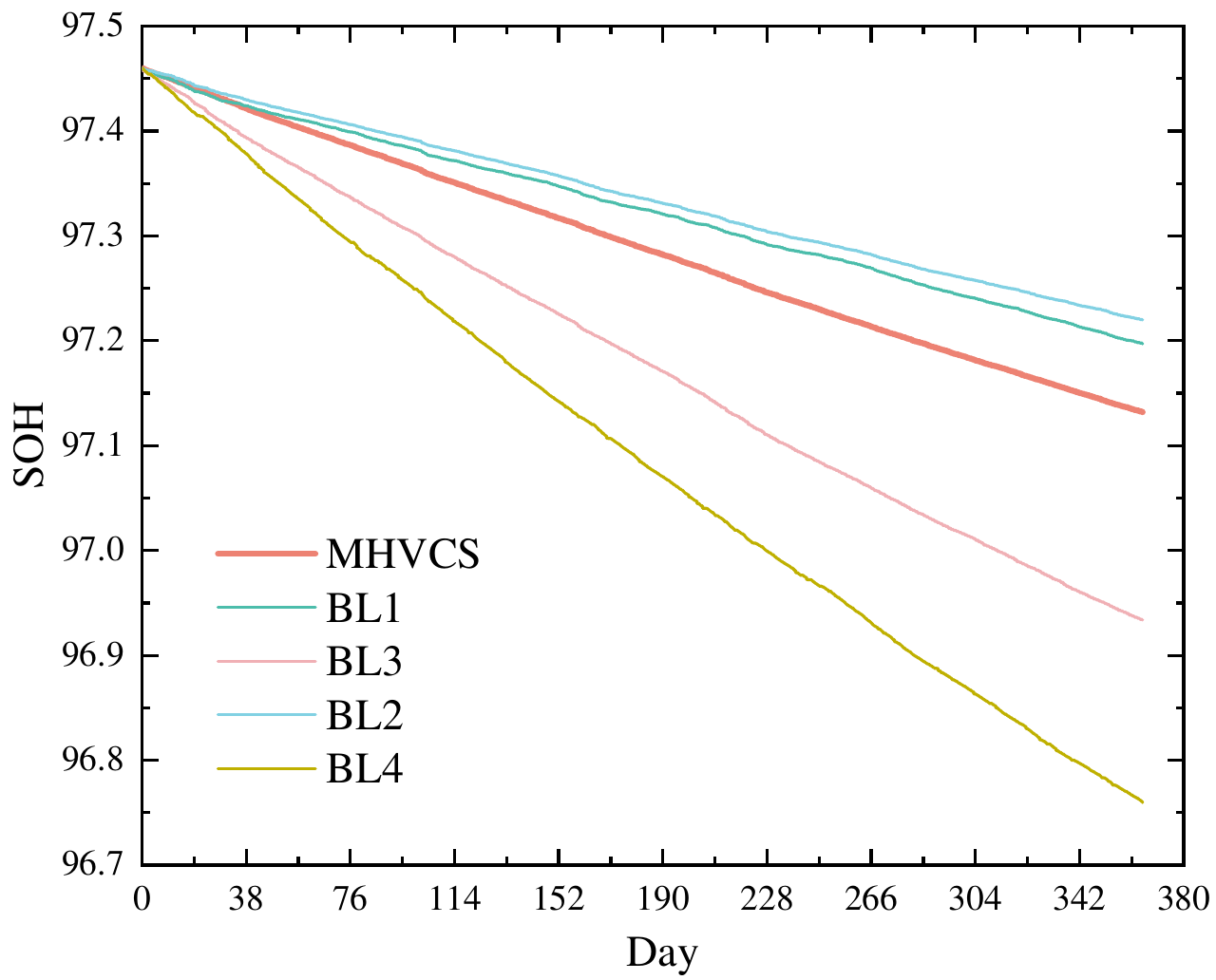}
\caption{SOH under MHVCS and four BLs strategies in one year.}
\label{fig:soh}
\end{figure}

The initial value of SOH was 97.46. After a year of strategy scheduling, the MHVCS has little difference in battery aging compared with the charging strategy BL1, BL2 and is better than the charging and discharging strategy BL3 and BL4.

The total cost $C$ and battery degradation cost $C_{bat}$ \cite{singh2020cost} are shown as:
\begin{align}
&C=C_{ch}+C_{bat}\\
&C_{bat}=\sum_{n=1}^{N}{(c_{bat}+\frac{c_{l}}{1-SOH_{min}}){(1-SOH^i_n)Q_n}}\label{cbattery}
\end{align}
The degradation cost of an EV battery is $c_{bat}=\$300/kWh$, and the labor cost for battery replacement is $c_{l}=\$240$. The EV battery is considered to be scrapped when the SOH drops below $80\%$. Hence the minimum SOH $SOH_{min}$ is $80\%$.

Evaluation indices (EIs) under different strategies are shown in Table~\ref{tab:soh}. EIs include one year's SOH after strategy optimization, load variance (LV), and a day's EV total cost.

\begin{table}[h]
 \centering
 \caption{\textsc{Evaluation indices under MHVCS and four BLs strategies.}}
 \label{tab:soh}
 \begin{tabular}{cccccc}
 \toprule
\makecell[c]{EIs} &\makecell[c]{MHVCS} &\makecell[c]{BL1} &\makecell[c]{BL2} &\makecell[c]{BL3} &\makecell[c]{BL4}\\ 
\midrule
SOH &97.13   & 97.20  &97.22   & 96.93 & 96.76\\
LV &71312.8  &508479.1   & 85010.7   & 67074.3 &485071.2\\ 
Cost &3127.8  &6602.6   & 3341.3   & 3267.2 &538.5\\
\bottomrule
 \end{tabular}
\end{table}

\begin{table}[h]
 \centering
 \caption{\textsc{Cost under MHVCS and four BLs strategies.}}
 \label{tab:cost}
 \begin{tabular}{cccccc}
 \toprule
\makecell[c]{Cost} &\makecell[c]{MHVCS} &\makecell[c]{BL1} &\makecell[c]{BL2} &\makecell[c]{BL3} &\makecell[c]{BL4}\\ 
\midrule
Charing Cost &1686.9  &5196.9   & 1945.7   & 1726.0 &-1088.1\\
Battery Cost &1440.9  &1405.7  & 1395.6   & 1541.2 &1626.6\\
\bottomrule
 \end{tabular}
\end{table}

Table \ref{tab:soh} and \ref{tab:cost} show that the proposed MHVCS can reduce the grid's load variance, the batteries' aging, and the total cost. The MHVCS yields a slightly lower SOH for battery aging than the charging strategies BL1 and BL2. However, it is significantly higher than the SOH obtained from discharging strategies BL3 and BL4. Regarding grid load stability, the LV resulting from our strategy is slightly larger than that obtained using the minimum grid load fluctuation strategy (BL3). However, it is significantly less than the LVs associated with scheduling strategies BL1, BL2, and BL4.
Regarding EV total costs, the optimized charging costs resulting from our strategy are only slightly higher than those obtained using the minimum charging cost strategy (BL4) while remaining lower than those associated with strategies BL1, BL2, and BL3. According to the EV cost analysis, the MHVCS is more economical than strategies BL1-3 but comparatively expensive compared to the minimum charging cost strategy (BL4). However, HCVS exhibits significantly lower grid load variance than BL4.

\begin{figure}[h]
\centering
\includegraphics[width=0.45\textwidth]{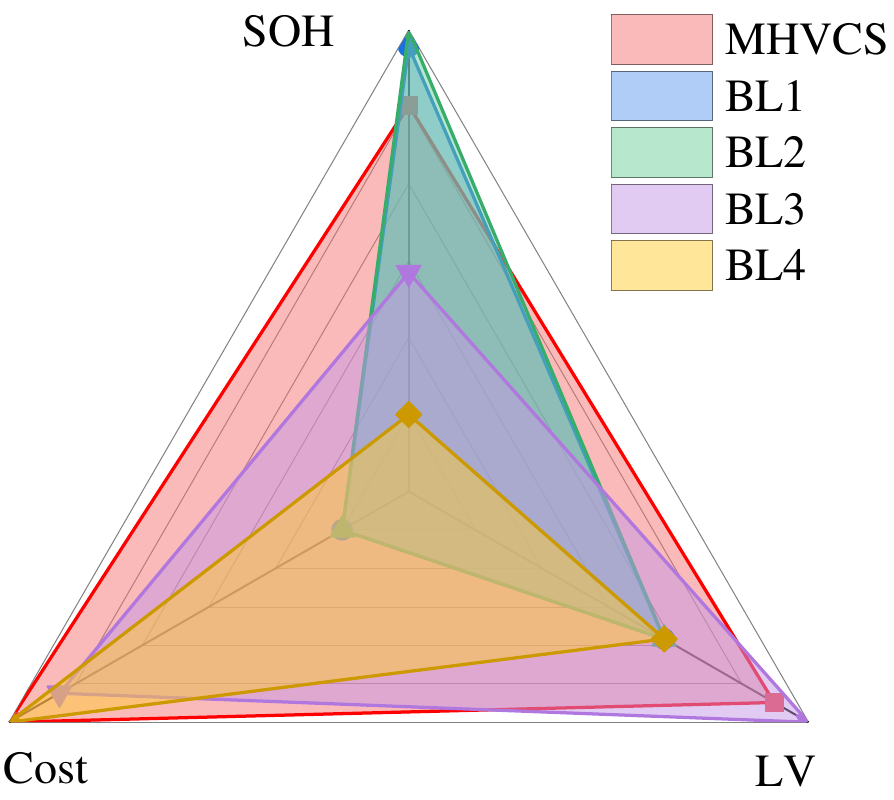}
\caption{Normalized SOH, LV, and cost evaluation indices under MHVCS and four BLs strategies.}
\label{fig:nei}
\end{figure}

To present the various EIs following strategy optimization more intuitively, we have utilized linear normalization to standardize the EIs. Larger normalized values indicate superior strategy performance concerning the corresponding EI. The normalized EIs under different strategies are shown in Fig.~\ref{fig:nei}.

From Fig.~\ref{fig:nei}, it is evident that MHVCS outperforms the other four BLs in the SOH, LV, and cost evaluation indices, exhibiting significantly higher numerical values. Therefore, the proposed MHVCS can achieve multi-stakeholder benefits in the scheduling strategy.

\section{Discussion}

SOC is the critical parameter to control the EV properly and to secure the power responses due to changes in operating conditions. The SOC constraints will ensure that the hierarchical coordination strategy satisfies the driving demand of EV users. The SOC constraints can reduce the long-term capacity fade rate and achieve a higher number of equivalent full cycles or a higher cumulative discharge capacity over the battery's useful life.

\begin{figure}[h]
    \centering
    \includegraphics[width=0.48\textwidth]{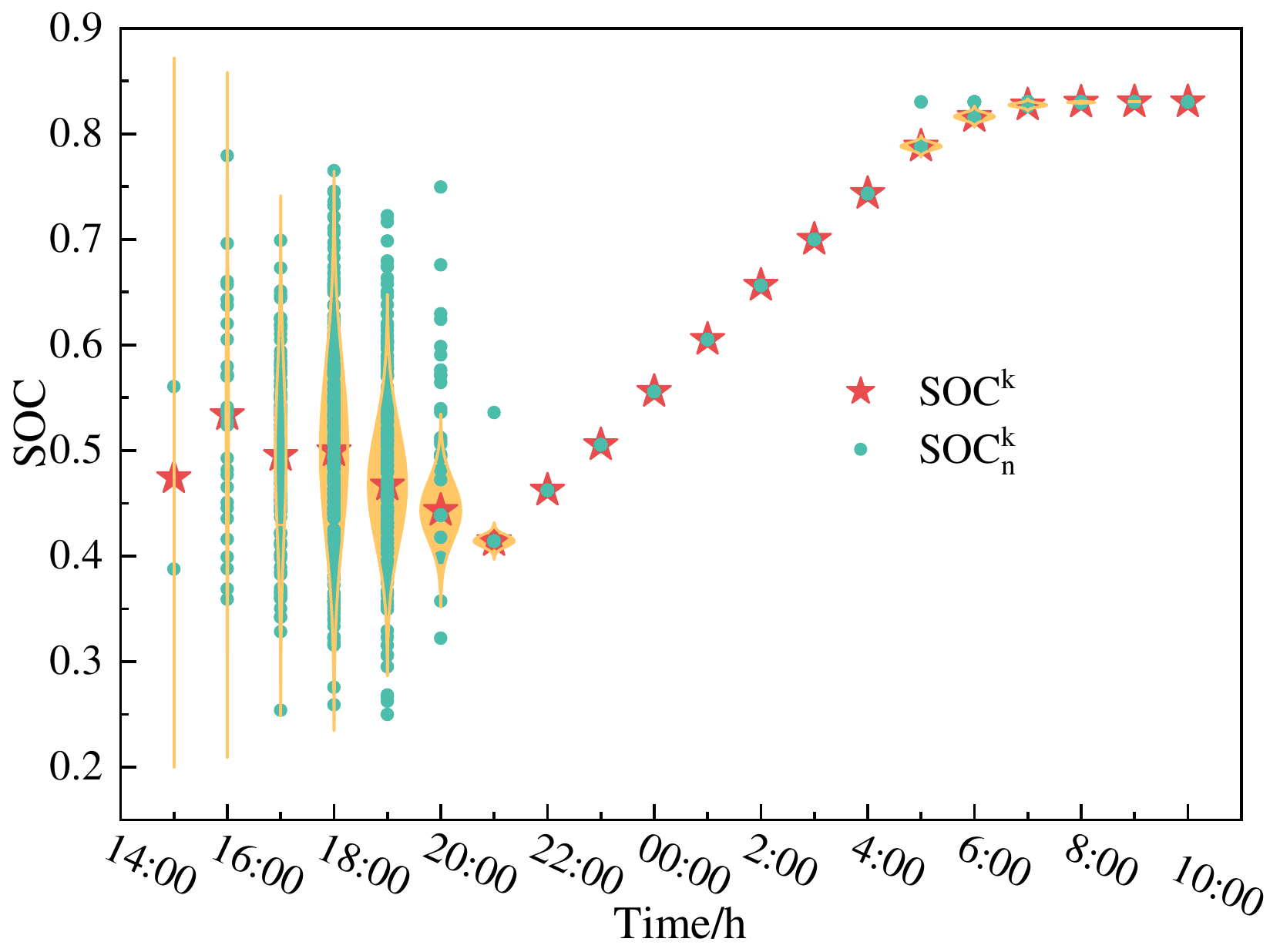}
		\caption{Distribution of individual EV SOC. }
		\label{fig:soc509pw}
\end{figure}

In Fig.~\ref{fig:soc509pw}, the red star indicates the SOC calculated based on the primary power strategy proposed by a selected validator in the context of the Proof of Stake algorithm. The scheduling period is from timeslot 15 to 10. At departure time, the SOC of EVA is 0.83, which meets the departure SOC needs (Between 0.8 and 0.9). When the EVA scheduling is completed, individual EV scheduling is required. The power allocation is elaborated in (\ref{kappa})-(\ref{soc1}). Therefore, the $n-th$ EV's SOC should always be satisfied (\ref{deps}). The EV SOCs in the PPO scheduling time from timeslot 15 to 19 are more dispersed than in the remaining time because EV SOCs differ a lot when EVs arrive home for the charging/discharging schedule. The scheduling is based on the EVA SOC. At the departure time, the SOC of the individual EV is 0.87. In the scheduling period, the SOC of individual EV is always between 0.2 and 0.9, satisfying the SOC constraint of the individual EV. The bounded SOC constraints prevent battery overcharge and over-discharge, protect the battery from rapid degradation, and extend the battery's service life.

\begin{figure}[h]
    \centering
    \includegraphics[width=0.45\textwidth]{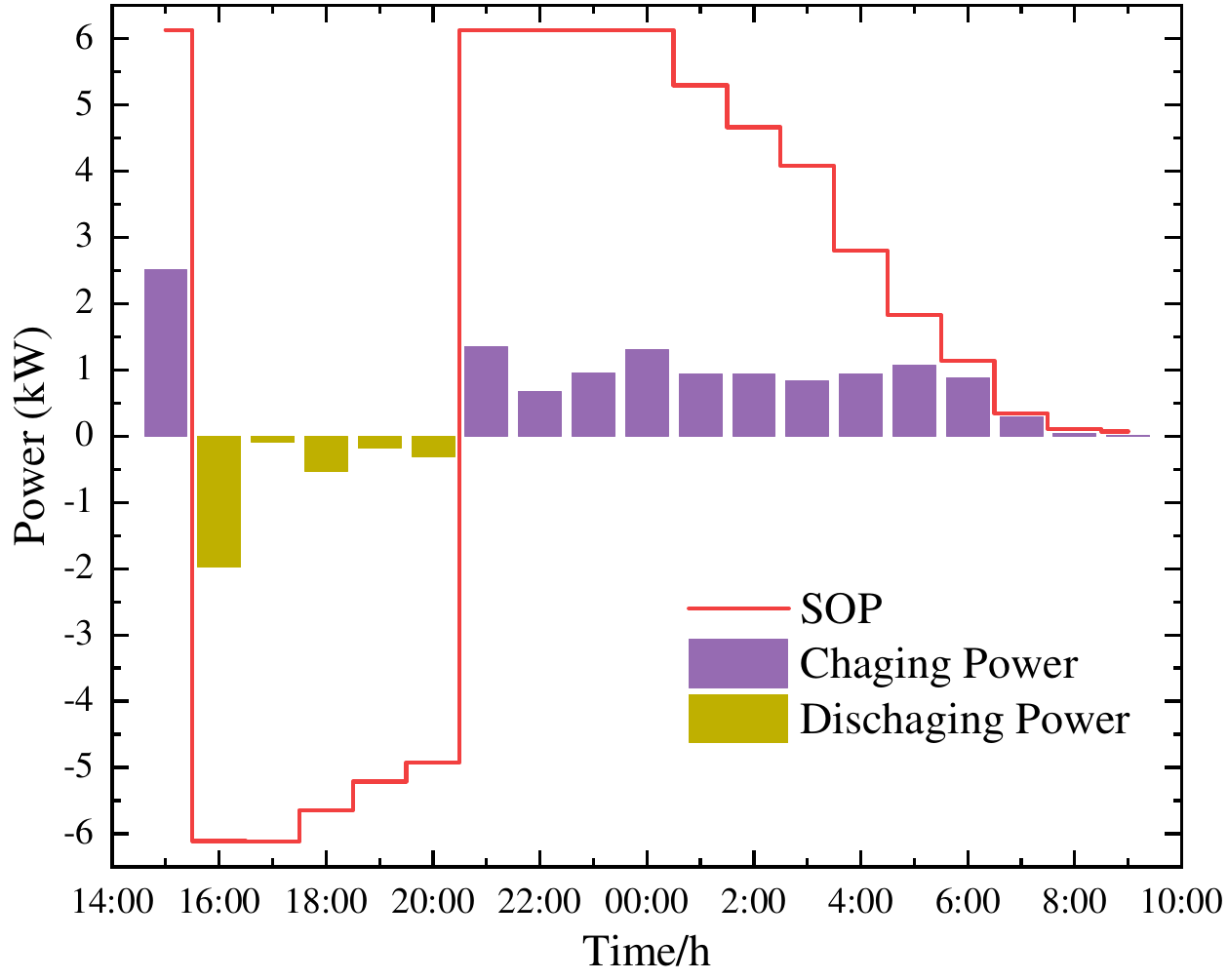}
    \caption{Single EV charging/discharging power in a day.}
    \label{fig:powersop}
\end{figure}

 Taking an EV during the scheduling period as a case study, the relationship between power and SOP is elucidated. As demonstrated in Fig.~\ref{fig:powersop}, throughout the scheduling period, the charging and discharging power constantly satisfies the SOP constraints of the EV battery.
  
\section{Conclusion}
This paper presents a MHVCS based on PPO and PoS algorithms for renewable energy integration and power grid stability. The proposed strategy considers the interests of the power grid, EVAs, and users, and utilizes a hierarchical scheduling framework. The comparative analysis demonstrates the outstanding, flexible, adaptable, and scalable performance of the proposed strategy for large-scale V2G under realistic operating conditions. The findings contribute to renewable energy and V2G research by providing a novel approach to V2G coordination that can optimize energy utilization, enhance power grid stability, and benefit all stakeholders involved. Further research can be conducted to explore additional applications of DRL in renewable energy integration and smart grid management.

\appendices
\section{Simulation Environment with Battery Conditioning}
Battery conditioning is a process that involves maintaining and improving the performance of batteries through a series of controlled charging and discharging cycles. It extends the battery's lifespan, improves capacity retention, and ensures optimal performance under varying operating conditions. Considering the actual states of the battery, we propose battery states (including the SOH, SOP, and SOC) constraints in the coordination control strategy.
\subsubsection{SOH Model}
Saxena et al.\cite{saxena2016cycle} established a capacity attenuation model for batteries under full or partial cycle conditions and obtained the relationship between average SOC and SOC variation and capacity loss rate. SOH is shown as:

\begin{align}
&SOH^i_n(\%)=100 - 3.25{SOC_{n,ave}^i}(1+3.25\Delta SOC_n^i\nonumber\\
&\qquad \qquad \qquad-2.25{\Delta SOC_n^i}^2)*(i/100)^{0.453}\\
&{SOC_{n,ave}^i}=0.5(\overline{SOC_n^i}+\underline{SOC_n^i})\\
&\Delta SOC_n^i=\overline{SOC_n^i}-\underline{SOC_n^i}\\
&i(m)=i(m-1)+\varepsilon (m-1) M_1\\
&\varepsilon (m)=\frac{0.5}{M(m-1)} (2-\frac{DOD(m-2)+DOD(m)}{DOD(m-1)})\nonumber\\
&\qquad \quad +\varepsilon (m-1)\\
&M(m)=H{{(\frac{DOD(m)}{100})}^{-\kappa }} {{I}_{dis,ave}}{{(m)}^{-{{\gamma }_{1}}}} {{I}_{ch,ave}}{{(m)}^{-{{\gamma }_{2}}}}
\end{align}
where $SOC_{n,ave}^i$ denotes the average SOC during the $i-{th}$ equivalent full cycle of the $n$-th EV. $\Delta SOC_n^i$ represents the variation in SOC uring the $i-{th}$ equivalent full cycle of the $n$-th EV. Additionally, $i$ signifies the equivalent full cycles endured by the battery. $\varepsilon$ is battery aging factor, $M_1$ is the the equivalent full cycle constant and is set to 1500 in the model. $M$ is the maximum number of cycles. $DOD$ stands for depth of discharge which refers to the percentage of the battery's capacity that has been discharged. $H$ is the cycle number constant, $\kappa$ is the exponent factor for the DOD, ${I}_{{dis,ave}/{ch,ave}}$ is the average discharge/charge current during a half cycle duration. ${\gamma }_{1}$ and ${\gamma }_{2}$ are exponent factor for the discharge/charge current, respectively. 

\subsubsection{SOP Model}
The battery SOP refers to the maximum charging and discharging power that the battery can withstand within a certain period of time. It is important for avoiding overcharging and over-discharging of the battery and extending its lifespan. In the model, the effects of SOC, battery terminal voltage and battery design are considered.

To prevent the SOC from exceeding the limit due to battery charging or discharging, the charging and discharging current of the battery is constrained based on the current SOC. The peak charging and discharging current of the battery during $k$ to $k+n$ time periods is set.
\begin{align}
&\overline{I^{k}_{SOC,dis}}=\cdot \cdot \cdot =\overline{I^{k+n}_{SOC,dis}}=\frac{Q_n(SOC^k-\underline{SOC})}{n{{T}_{s}}}\\
&\overline{I^{k}_{SOC,ch}}=\cdot \cdot \cdot =\overline{I^{k+n}_{SOC,ch}}=\frac{Q_n(\overline{SOC}-SOC^k)}{n{{T}_{s}}}
\end{align}
where, $\overline{I^{k}_{SOC,dis}}$ is the maximum instantaneous discharge current based on battery SOC constraints at timeslot $k$. $\overline{I^{k}_{SOC,ch}}$ is the maximum instantaneous charging current based on battery SOC constraints at timeslot $k$. ${{T}_{s}}$ is the sampling interval. ${{Q}_{n}}$ is the current maximum available capacity of the lithium battery.

The battery voltage at the terminal cannot exceed the design limit during operation, so the battery terminal voltage is constrained. The peak charging and discharging current of the battery during $k$ to $k+n$ time periods is set.
\begin{align}
&\overline{I^{k}_{{U_T},dis}}=\cdot \cdot \cdot =\overline{I^{k+n}_{{U_T},dis}}=\frac{{{U}_{oc}}(SOC^{k})-\underline{U_{T}}}{{{R}_{0}}+\frac{n{{T}_{s}}}{{{Q}_{n}}}* \frac{\partial {U_{OC}}(SOC_{k})}{\partial SOC_{k}}}\\
&\overline{I^{k}_{{U_T},ch}}=\cdot \cdot \cdot =\overline{I^{k+n}_{{U_T},ch}}=\left| \frac{{{U}_{oc}}(SOC^k)-\overline{U_T}}{R_0+\frac{n{T_s}}{Q_n}* \frac{\partial {U_{OC}}(SOC^k)}{\partial SOC^k}} \right|
\end{align}
where $\overline{I^{k}_{{U_T},dis}}$ is the maximum instantaneous discharge current based on battery terminal voltage constraints at timeslot $k$.
$\overline{I^{k}_{{U_T},ch}}$ is the maximum instantaneous charging current based on battery terminal voltage constraints at timeslot $k$.
${U_{oc}}(SOC^k)$ is the battery terminal voltage at timeslot $k$.
${R_0}$ is the internal resistance of the battery.
$\overline{U_T}$ and $\underline{U_T}$ are the limits of the battery terminal voltage.

The peak current of the lithium battery under multiple constrained factors.
\begin{align}
&\overline{I^{k}_{dis}}=\min \{ \overline{I^{k}_{SOC,dis}},\overline{I^{k}_{{U_T},dis}},\overline{I^{k}_{{design},dis}}\}\\
&\overline{I^{k}_{ch}}=\min \{ \overline{I^{k}_{SOC,ch}},\overline{I^{k}_{{U_T},ch}},\overline{I^{k}_{{design},ch}}\}
\end{align}
where $\overline{I^{k}_{dis}}$ is the maximum discharge current of the battery at timeslot $k$.
$\overline{I^{k}_{ch}}$ is the maximum charging current of the battery at timeslot $k$.
$\overline{I^{k}_{{design},dis}}$ is the maximum discharge current of the battery design.
$\overline{I^{k}_{{design},ch}}$ is the maximum charging current of the battery design.

The $\overline{I^{k}_{dis}}$ and $\overline{I^{k}_{ch}}$ can be used to obtain the continuous peak charging and discharging power of the battery.
\begin{align}
&\overline{P_{k}^{dis}}={U_T^k}* \overline{I^{k}_{dis}}\\
&\overline{P_{k}^{ch}}={U_T^k}* \overline{I^{k}_{ch}}
\end{align}
where $\overline{P_{k}^{dis}}$ is the continuous peak discharge power of the battery under multiple constraints at timeslot $k$.
$\overline{P_{k}^{ch}}$ is the continuous peak charging power of the battery under multiple constraints at timeslot $k$.
${U_T^k}$ is the battery terminal voltage at timeslot $k$.

\ifCLASSOPTIONcaptionsoff
  \newpage
\fi

\bibliographystyle{IEEEtran}
\bibliography{ref}

\end{document}